\documentclass{article}
\usepackage{spconf}
\usepackage{amsmath,graphicx,amssymb}
\usepackage{xcolor}
\usepackage{siunitx}
\usepackage{lipsum}
\usepackage{booktabs}
\usepackage{scalerel}
\PassOptionsToPackage{hyphens}{url}
\usepackage{hyperref}
\usepackage{url}
\usepackage{breakurl}
\usepackage{caption}
\usepackage{float}
\usepackage{multirow}
\usepackage{cite}
\usepackage{subcaption}

\def\x{{\mathbf x}}
\def\L{{\cal L}}

\def\coarse{{\mathcal C}}
\def\fine{{\mathcal F}}

\title{LMCODEC: A LOW BITRATE SPEECH CODEC WITH CAUSAL TRANSFORMER MODELS}
\name{\parbox{\linewidth}{\centering Teerapat Jenrungrot\textsuperscript{1}\thanks{This work was done during a research internship at Google.}, Michael Chinen\textsuperscript{2}, W. Bastiaan Kleijn\textsuperscript{2,3}, Jan Skoglund\textsuperscript{2}, \\
\textit{Zalán Borsos\textsuperscript{2}, Neil Zeghidour\textsuperscript{2}, Marco Tagliasacchi\textsuperscript{2}}}}

\address{
\textsuperscript{1}University of Washington, Seattle \\
\textsuperscript{2}Google \\
\textsuperscript{3}School of Engineering and Computer Science, Victoria University of Wellington
}

\expandafter\def\expandafter\UrlBreaks\expandafter{\UrlBreaks% save the current one
  \do\a\do\b\do\c\do\d\do\e\do\f\do\g\do\h\do\i\do\j%
  \do\k\do\l\do\m\do\n\do\o\do\p\do\q\do\r\do\s\do\t%
  \do\u\do\v\do\w\do\x\do\y\do\z\do\A\do\B\do\C\do\D%
  \do\E\do\F\do\G\do\H\do\I\do\J\do\K\do\L\do\M\do\N%
  \do\O\do\P\do\Q\do\R\do\S\do\T\do\U\do\V\do\W\do\X%
  \do\Y\do\Z\do\*\do\-\do\~\do\'\do\"\do\-}

\begin{document}
\newcommand{\note}[1]{\textcolor{blue}{#1}}
\ninept
\maketitle
\linepenalty=10000000
% \begin{abstract}
% We present a fully causal speech coding architecture that provides high quality at 1000 bits per second.  The method uses causal transformer-based models on the discrete residual codebook tokens from a pretrained SoundStream model that is trained for speech coding at 6 kilobits per second.  One transformer model predicts the fine SoundStream tokens from the coarse tokens, allowing transmission of fewer codes.  Another transformer model predicts the uncertainty for the next frame given the past frames, and is used to conditionally entropy code each token.  A MUSHRA subjective test was conducted and shows that the quality is comparable to reference codecs at higher bitrates. Example audio is available at \url{https://sites.google.com/corp/google.com/lowbitrateaudiolm-results}.
% \end{abstract}
\begin{abstract}
We introduce LMCodec, a causal neural speech codec that provides high quality audio at very low bitrates. The backbone of the system is a causal convolutional codec that encodes audio into a hierarchy of coarse-to-fine tokens using residual vector quantization. LMCodec trains a Transformer language model to predict the fine tokens from the coarse ones in a generative fashion, allowing for the transmission of fewer codes. A second Transformer predicts the uncertainty of the next codes given the past transmitted codes, and is used to perform conditional entropy coding. A MUSHRA subjective test was conducted and shows that the quality is comparable to reference codecs at higher bitrates. Example audio is available at  \url{https://mjenrungrot.github.io/chrome-media-audio-papers/publications/lmcodec}.
\end{abstract}

\section{Introduction}
\label{sec:intro}

Speech coding, which consists of compressing speech signals to a limited number of bits with minimal distortion, is at the core of communication technologies such as mobile telephony or Voice over IP (VoIP). Opus~\cite{valin2012opus} and EVS~\cite{dietz2015evs} are state-of-the-art speech coding techniques that combine traditional coding tools, such as Linear Predictive Coding (LPC), Code Excited Linear Prediction (CELP), and Modified Discrete Cosine Transformation (MDCT) to achieve high coding efficiency over different content types and bitrates. These waveform and parametric codecs rely on psychoacoustics expertise to design signal processing pipelines with maximal coding efficiency. Yet, while fast and interpretable, such handcrafted pipelines \linebreak only represent a fraction of the potential models for a speech codec.

This has motivated data-driven approaches to train neural networks to perform speech coding. These networks leverage large amounts of training data while relaxing the assumptions made on the type of transformations applied by the system \cite{morishima1990speechcoding, kankanahalli2018speechdnn, valin2019lpcnet, garbacea2019vqvae, polyak2021speech, zhen2019cascaded, petermann2021harp, zeghidour2021soundstream}. In particular, the SoundStream neural codec combines a causal convolutional architecture with a residual vector quantizer. This quantization method produces a hierarchy of coarse-to-fine codes, and allows for efficient compression while providing bitrate scalability. As a result, SoundStream at \SI{3}{~}kbps matches the quality Opus at \SI{12}{~}kbps. However, the quality of most codecs, be they handcrafted or trained, degrades significantly at bitrates lower than \SI{3}{~}kbps.

In this work, we introduce LMCodec, a low bitrate speech codec that combines recent advances in neural audio coding and audio generative modeling. LMCodec uses autoregressive Transformers \cite{attentionvaswani} on SoundStream tokens to (i) model the entropy of the distribution of coarse tokens and (ii) predict fine tokens from the coarse ones. At inference, LMCodec extracts the codes of a  SoundStream model from the input waveform. However, instead of sending all codes to the receiver like a SoundStream codec would do, LMCodec only transmits entropy-coded coarse tokens. On the receiver side, a generative language model is used to predict fine tokens from the coarse ones, and a SoundStream decoder then reconstructs audio from the complete token sequence.

LMCodec takes inspiration from the AudioLM \cite{borsos2022audiolm} generative model, which also predicts fine SoundStream tokens from coarse ones. However, unlike AudioLM, LMCodec does low bitrate compression rather than generative modeling, and to do so leverages AudioLM both as a generative model and an entropy model. Other Transformer-based models for low bitrate coding have been proposed ~\cite{siahkoohi2022ultra, polyak2021speech}. The codec in~\cite{siahkoohi2022ultra} enriches SoundStream with embeddings extracted from a self-supervised speech representation model \cite{conformer} and achieves speech compression at a rate of \SI{600}{~}bps. \cite{polyak2021speech} synthesizes speech from a combination of phonetic, pitch and speaker representations to achieve 365 bps. Unlike these models, LMCodec is a fully causal model, which is thus amenable to online encoding and decoding. Our primary contribution is the design of a new neural speech codec, which achieves state-of-the-art results  outperforming many previous codecs operating at three to four times the rates according to subjective human evaluation metrics.

Subjective evaluations demonstrate how LMCodec allows for low bitrate speech coding with minimal distortion, with LMCodec at approximately \SI{1}{}-\SI{1.5}{~}kbps matching the performance of Opus at \SI{12}{~}kbps. We furthermore analyze the failure modes of our system, as well as the discrepancies in bit allocations between speech and non-speech sections of an audio signal.

\begin{figure}
    \centering
    \captionsetup{belowskip=-20pt}
    \centerline{\includegraphics[trim={0 3cm 0 1cm},clip,width=8.5cm]{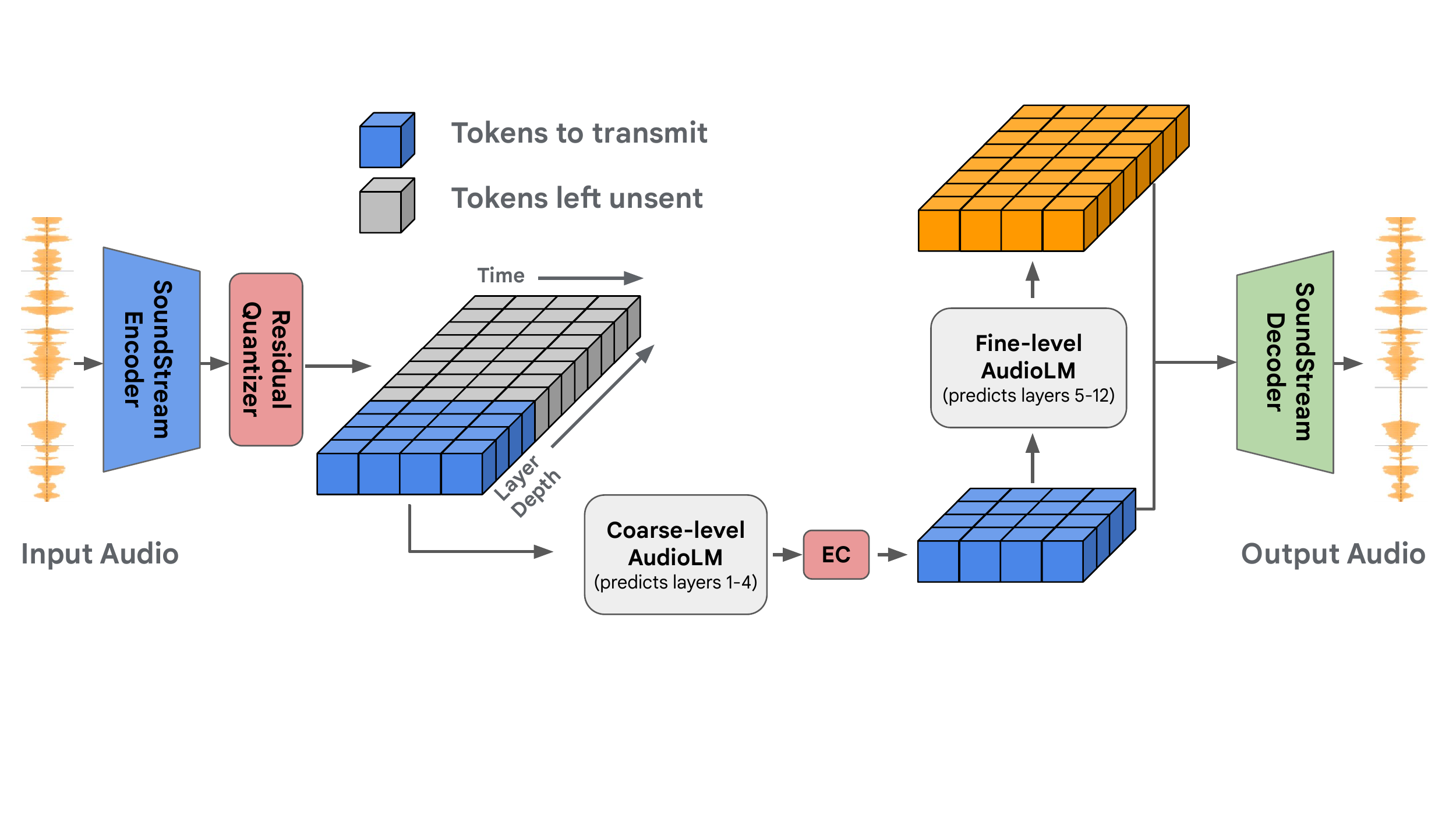}}
    \caption{Overall pipeline of the proposed codec.}
    \label{fig:overall}
\end{figure}

\begin{figure}
    \captionsetup{belowskip=-10pt}
    \centering
    \begin{subfigure}[b]{0.245\textwidth}
         \centering
         \includegraphics[clip, trim=0.5cm 0.5cm 1.3cm 1.7cm, height=3.9cm]{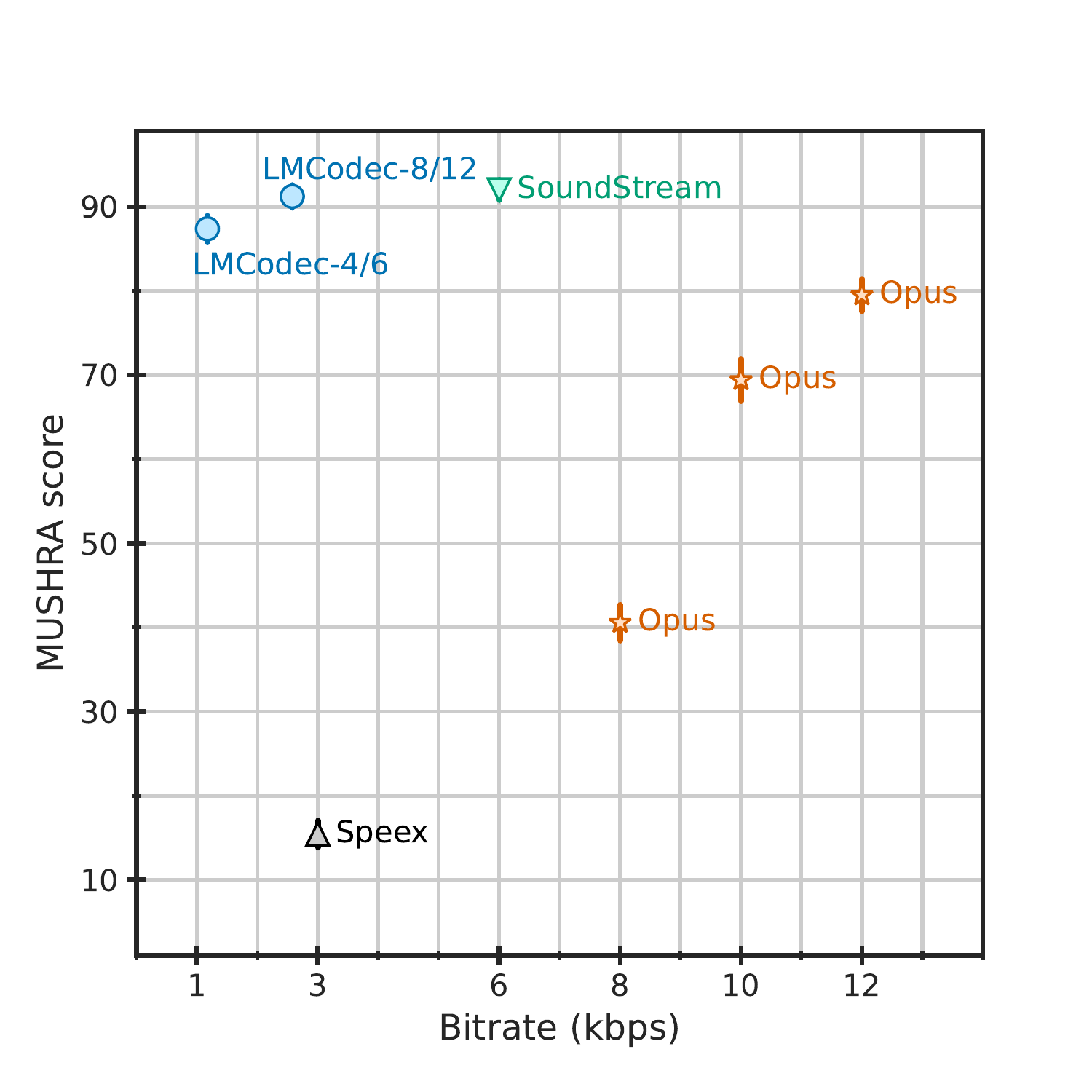}
         \label{fig:a}
    \end{subfigure}
    \kern-1em
    \begin{subfigure}[b]{0.245\textwidth}
        \centering
         \includegraphics[clip, trim=0.5cm 0.5cm 1.3cm 1.7cm, height=3.9cm]{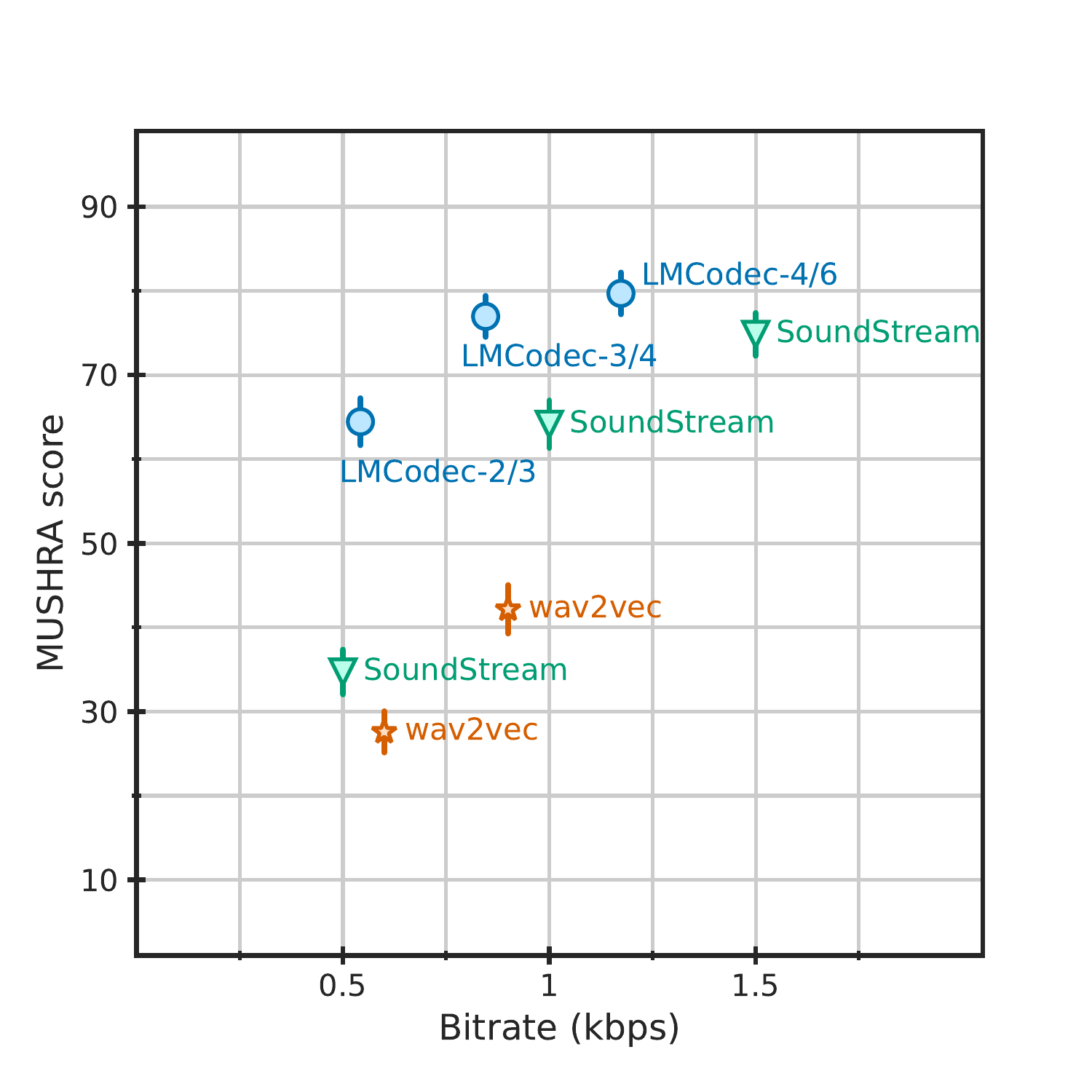}
         \label{fig:b}
    \end{subfigure}
    \caption{MUHSRA-like subjective evaluation from state-of-the-art codecs with medium and low bitrates. LMCodec-$x/y$ refers to our model with $N_\mathcal{C}=x$ and $N_\mathcal{C} + N_\mathcal{F}=y$. wav2vec \cite{siahkoohi2022ultra} is a recent neural codec based on SoundStream and Transformer.}
    \label{fig:mushra} \label{fig:mushra_new}
\end{figure}

\section{Proposed Model}
\label{sec:model}

In this section, we describe our proposed speech codec consisting of four components: an encoder, a residual quantizer, an AudioLM block, and a decoder. The encoder, residual quantizer, and decoder follow similar structures from SoundStream. 
% Please refer to~\cite{zeghidour2021soundstream} for complete details on SoundStream. 
At the very high level, the encoder takes raw speech in the time domain as an input and extracts low-rate features that contain sufficient information to reconstruct the speech. The residual quantizer finds discrete representations of the inherently continuous encoded features. AudioLM poses the modeling of the quantized discrete representation as a language modeling problem and estimates the probability distribution of the next discrete audio token given previous audio tokens. Finally, the decoder reconstructs the input speech signal from the discrete encoded features.

\subsection{SoundStream}

We now briefly describe the SoundStream model \cite{zeghidour2021soundstream} that we used for creating high-quality audio tokens.

\subsubsection{Encoder}
Given a raw speech signal $\mathbf{x} \in [-1,1]^T$ of length $T$, the encoder $\mathcal{E}: [-1,1]^T \rightarrow \mathbb{R}^{T_e \times N_e}$ creates a sequence of embeddings of length $T_e \ll T$, each with dimension $N_e$. In our proposed model, the encoder takes raw waveform speech at $T = \SI{16}{kHz}$ as input and generates $N_e=128$ dimensional speech features with a frame rate of $\SI{50}{Hz}$. The architecture of the encoder is fully convolutional based on causal 1D convolutions. Hence, the algorithmic delay is determined by the overall striding factor (i.e., $T/T_e = 320$ samples or \SI{20}{ms}).

\subsubsection{Residual Vector Quantizer (RVQ)}

Transmission of continuous speech features over low-bandwidth channels is achieved via vector quantizers (VQs) \cite{zeghidour2021soundstream}, where the features are turned into discrete representations while introducing minimal distortion. Given the encoded features $\mathbf{e} \in \mathbb{R}^{T_e \times N_e}$, the residual quantizer $\mathcal{Q}: \mathbb{R}^{T_e \times N_e} \rightarrow \{0, \ldots, 2^{\lceil\log N_c\rceil}\scalebox{1.75}[1.0]{\text{-}}1\}^{T_e \times N_q}$ computes the corresponding binary representation of $\mathbf{e}$ and its inversion, where $N_q$ is the number of quantizers and $N_c$ is the codebook size of a single quantizer. In our proposed model, we always use the codebook of size $N_c = 2^{10}$ and vary the number of layers in the residual VQs: $N_q \in \{3,4,6,12,24\}$.

\subsubsection{Decoder}

The decoder $\mathcal{D}\!:\!\, \mathbb{R}^{T_e \times N_e} \rightarrow [-1,1]^T$ synthesizes the original speech signal from the post-quantized embeddings. In our work, we adopt the CNN-based decoder method trained with adversarial loss in addition to losses on waveform and spectral domains. The architecture of the decoder is similar to that of the encoder, with a transposed convolutional layer to upsample the output. The adversarial training framework relies on two types of discriminators: waveform domain and short time Fourier Transform (STFT) domain discriminators.

\subsection{AudioLM}

\begin{figure}
    \captionsetup{belowskip=-10pt}
    \centering
    \begin{subfigure}[b]{0.26\textwidth}
         \centering
         \includegraphics[clip, trim=0 0 0 0,height=3.9cm]{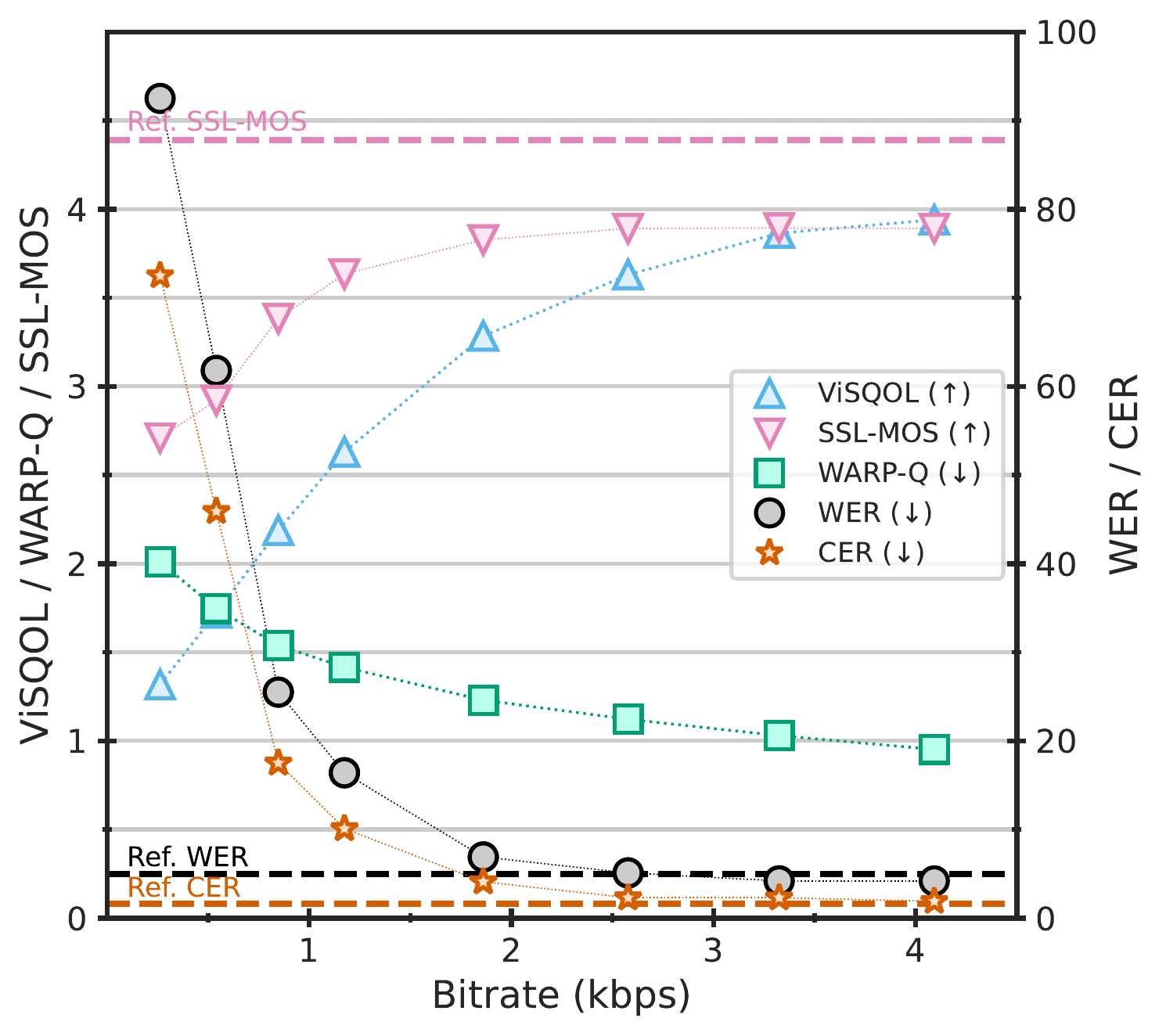}
    \end{subfigure}
    \begin{subfigure}[b]{0.21\textwidth}
        \centering
        \includegraphics[clip, trim=0 0 0 0, height=3.9cm]{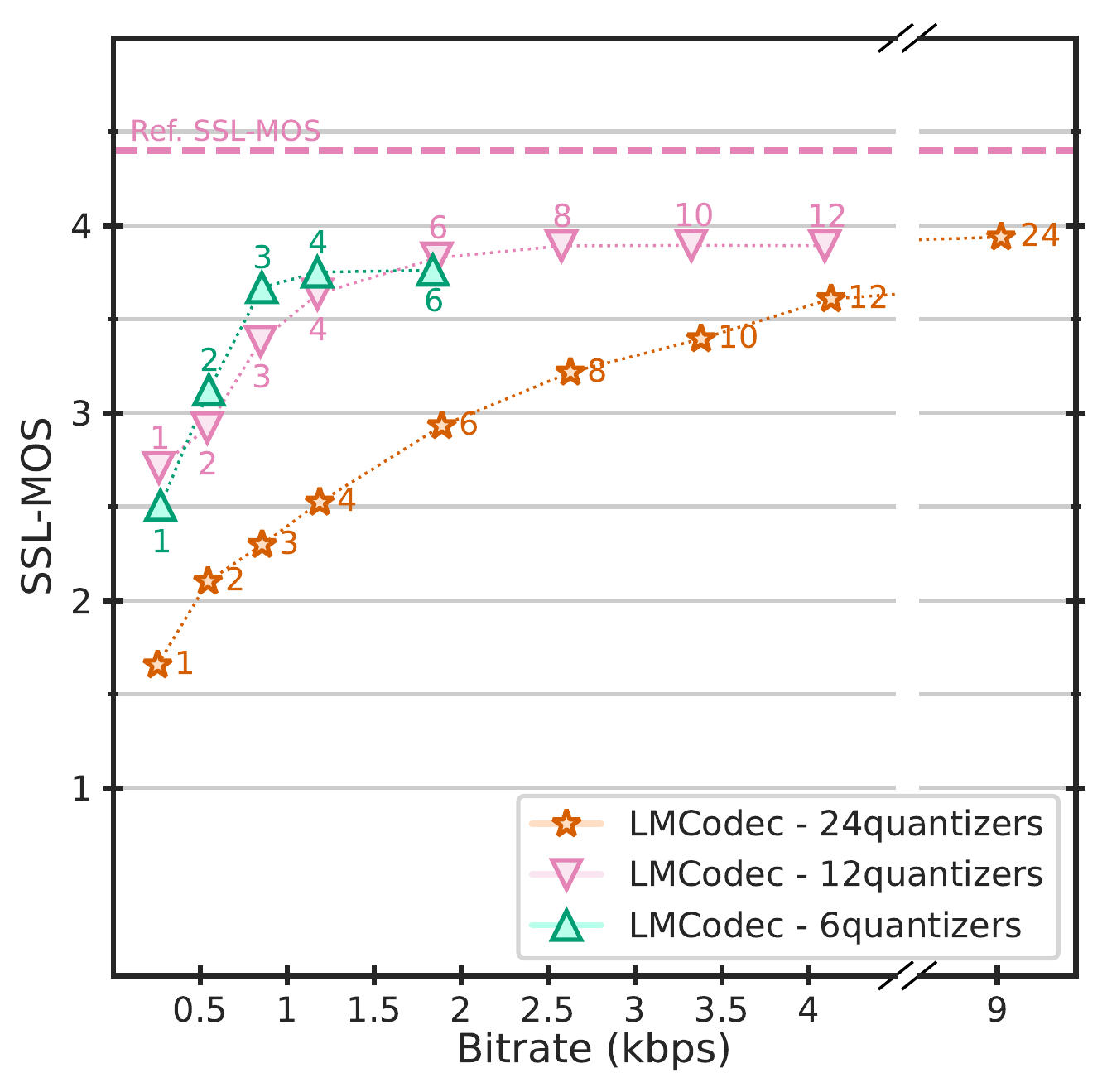}
    \end{subfigure}
    \caption{Objective evaluation of different LMCodec models. (left) LMCodec with a fixed number of RVQ layers (i.e., $N_\mathcal{C}+N_\mathcal{F} = 12$) on various standard metrics. (right) LMCodec with $N_\mathcal{C}+N_\mathcal{F} \in \{6,12,24\}$ on SSL-MOS \cite{cooper2022generalization}. Numbers next to the markers refer to the number of coarse-level codes $N_\mathcal{C}$.}
    \label{fig:objective_eval}
\end{figure}

In this subsection, we describe the problem of language modeling of SoundStream tokens. Adding a language model in the bottleneck enables interesting modeling tasks, including modeling the distribution of future SoundStream tokens (Section \ref{sec:coarse-audiolm}) or tokens at different VQ layers (Section \ref{sec:fine-audiolm}).

For the rest of this paper, let $N_{\coarse}$ and $N_{\fine}$ denote the number of quantizers for the coarse-level and fine-level AudioLMs, respectively. Figure \ref{fig:overall} shows the overall architecture of our proposed model, in which we use $N_{\coarse} = 4$ and $N_{\fine} = 8$. In our experiment, we use various combination of $(N_\mathcal{C}, N_\mathcal{F})$ ranging from $N_\mathcal{C} + N_\mathcal{F} = 3$ to $N_\mathcal{C} + N_\mathcal{F} = 24$. Additionally, let $c^{(n)}_{k}$ denote the SoundStream token at frame $n$ and VQ layer $k$.

\subsubsection{Coarse-level AudioLM\label{sec:coarse-audiolm}}

The goal of the coarse-level AudioLM is to model the distribution of the next coarse SoundStream tokens. Specifically, we are interested in modeling the conditional distribution of the next SoundStream tokens given the past information 
\begin{equation}
p_{\coarse}\Big(c^{(n)}_{k} \Bigm| \underbrace{c^{(n)}_{k-1}, \ldots, c^{(n)}_{1}}_{\text{coarse-level current frame}}, \underbrace{c^{(n-1)}_{N_\coarse}, \ldots, c^{(1)}_{1}}_{\text{past information}}\Big)
\end{equation}
for $k \in \{1, \ldots, N_\coarse\}$.

Given the distribution of the future SoundStream tokens, we build a codec by using lossless Entropy Coding (Section \ref{sec:entropy_coding}). More specifically, the discrete probability distribution of SoundStream tokens can be estimated both at the sender and the receiver sides, and we use this to drive an entropy codec. Note that in our proposed method, we only need to transmit $N_{\coarse}$ tokens per single audio frame. The remaining $N_{\fine}$ tokens are generated at the receiver side only as described in the next section.

\subsubsection{Fine-level AudioLM\label{sec:fine-audiolm}}
Similar to the coarse-level AudioLM, the fine-level AudioLM predicts the top VQ layers given the information about bottom VQ layers in addition to the past information. Specifically, we are interested in modeling the distribution of the fine-level SoundStream tokens conditioned on the coarse-level tokens and the past information:
\begin{equation}
\resizebox{.9\hsize}{!}{$p_{\fine}\Big(c^{(n)}_{k} \Bigm| \underbrace{c^{(n)}_{k-1}, \ldots, c^{(n)}_{N_\coarse+1}}_{\text{fine-level current frame}}, \underbrace{c^{(n)}_{N_\coarse}, \ldots, c^{(n)}_{1}}_{\text{coarse-level current frame}}, \underbrace{c^{(n-1)}_{N_\coarse + N_\fine}, \ldots, {c}^{(1)}_{1}}_{\text{past information}}\Big)$}
\end{equation}
for $k \in \{N_{\coarse}+1, \ldots, N_{\coarse} + N_{\fine}\}$. Note that our model is causal, in contrast to AudioLM.

Since we only transmit the coarse-level tokens, we model the distribution of the fine-level tokens by assuming that we have access to ground-truth coarse-level SoundStream tokens. We note that, while \cite{borsos2022audiolm} also proposes a similar fine-level AudioLM stage, our contribution here is the causal formulation of the task, which makes our approach more suitable and amenable to online decoding.

\subsection{Entropy Coding (EC)\label{sec:entropy_coding}}

Given the distribution of coarse-level SoundStream tokens, we transmit data by using entropy coding, a lossless data compression technique. In this work, we provide experimental results using Huffman coding, in addition to the estimated entropy rate. We treat each code from the residual VQs separately and do not perform any grouping to reduce the upper bound on the bitrate.

\begin{figure}
    \centering
    \captionsetup{belowskip=-13pt}
    \centerline{\includegraphics[clip, trim=3.5cm 0.5cm 3.2cm 2.2cm, width=8cm]{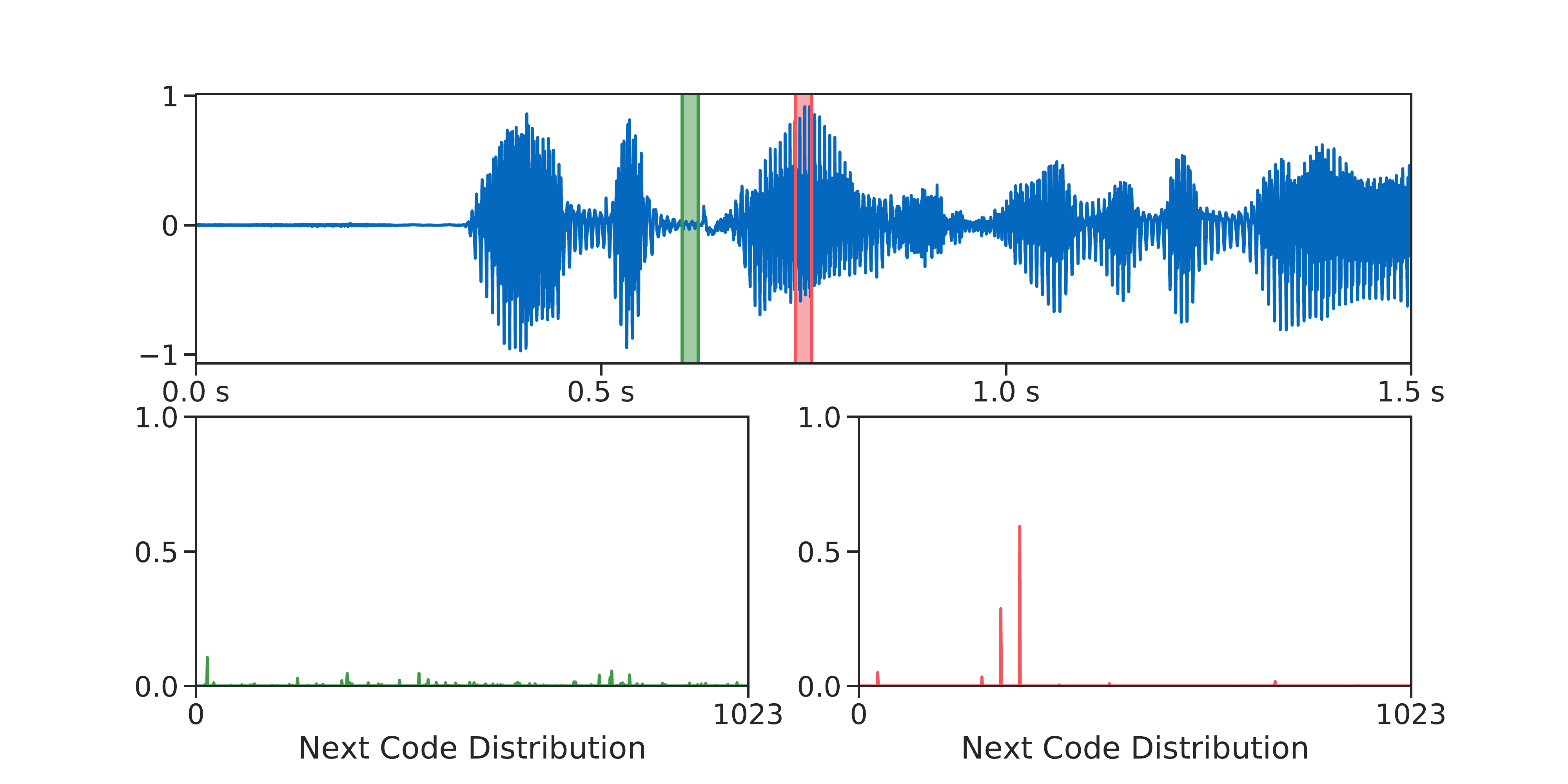}}
    \caption{Distribution of codes prediction for inputs from the non-voice section and inputs from the middle of phonemes }
    \label{fig:audio-output}
\end{figure}

We first note that our proposed codec requires only sending coarse-level SoundStream tokens using entropy coding. Specifically, given raw audio, LMCodec first encodes audio into SoundStream tokens and models the probability distribution of the next SoundStream tokens, driving the entropy codec. 
Note that the discrete probability distribution of SoundStream tokens can be estimated both at the sender and the receiver sides, so the receiver can losslessly reconstruct the coarse tokens. 
To generate audio output from only coarse-level tokens, we use a fine-level AudioLM to synthesize fine-level tokens from the transmitted coarse-level tokens and then generate audio from both coarse-level and fine-level tokens using \linebreak SoundStream decoder.

\subsection{Training Strategy}

We adopt a 2-stage training paradigm. First, we train only the encoder, quantizer, and decoder. Then, we freeze the weights of these components and train only the AudioLM components. We train the coarse-level and fine-level AudioLM models separately.

\subsubsection{Loss Functions}

We trained the SoundStream model using the standard adversarial loss, feature matching loss, reconstruction loss, and quantization loss according to \cite{zeghidour2021soundstream}. In training AudioLM models, we use the standard cross-entropy loss for language modeling over the vocabulary space.

\subsubsection{Training configurations}

To create our codec modules, we adapted the architectures of the encoder, quantizer, generator, and discriminators used in SoundStream \cite{zeghidour2021soundstream} and AudioLM from \texttt{T5X}. Both AudioLM models are the decoder-only models based on the \texttt{base} model of \texttt{t5.1.1} (with approximately 250 million parameters).
% pre-trained on multilingual C4 dataset \cite{raffel2020exploring}.

The SoundStream model is trained on \SI{16}{kHz} audio from the LibriVox dataset \cite{kearns2014librivox} for 1M steps.
% using a batch size of \textcolor{red}{???} and \textcolor{red}{optimizer}. 
Both coarse-level and fine-level AudioLM models are trained on \SI{16}{kHz} audio from the Libri-Light dataset \cite{kahn2020libri} for 1M steps with a batch size of 32 and sequence length of 1024 SoundStream tokens with Adafactor optimizer \cite{shazeer2018adafactor} with a decay rate of 0.8. 

We trained multiple coarse-level and fine-level AudioLM models to achieve varieties of bitrates. The bitrates are calculated based on the entropy coding of codes from coarse-level AudioLM.

\section{Evaluation}
\label{sec:evaluation}

To demonstrate the performance of our proposed method, we evaluate LMCodec using both objective and subjective evaluations. For objective evaluation, we report the accuracy of LMCodec future token prediction and objective metrics including ViSQOL \cite{chinen2020visqol}, WARP-Q \cite{jassim2021warp}, SSL-MOS \cite{cooper2022generalization}, WER, and CER together with bitrate based on the test split from the clean LibriSpeech dataset \cite{panayotov2015librispeech}.

For subjective evaluation, we perform two MUSHRA-like \cite{recommendation20151534} subjective tests to compare the audio quality with standard state-of-the-art speech codecs at medium bitrate (i.e., \SI{1}{~}kbps to \SI{12}{~}kbps) and low rate (i.e., \SI{0.5}{~}kbps to \SI{1.5}{~}kbps). The tests were conducted respectively on 91 and 94  crowd-sourced raters using headphones over 32 clean utterances from VCTK dataset \cite{Yamagishi2019CSTRVC}. Raters who did not score the reference above 80 at least 80\% of the time were discarded, as were raters who rated more than 75\% of non-reference samples 80 or above. 40  raters for the medium rate test and 33 raters for the low rate test met this requirement.

As shown in Figure \ref{fig:mushra}, the raters found that LMCodec-4/6 with 4~quantizers at \SI{1.1}{~}kbps perform significantly better than \SI{12}{~}kbps Opus. LMCodec-8/12 with 8~quantizers at \SI{2.6}{~}kbps has comparable performancce to SoundStream at \SI{6}{~}kbps. The low-rate MUSHRA test compares recent transformer neural codecs and lower bitrate SoundStream models.  The raters preferred LMCodec to the transformer models from~\cite{siahkoohi2022ultra} and SoundStream at the same rate.

\subsection{Discussion}

Table \ref{tab:accuracy} shows the accuracy of the future token prediction and the bitrate performance of LMCodec from the test split of the clean LibriSpeech \cite{panayotov2015librispeech}. For accuracy, we note that perfect accuracy means the model knows perfectly what the next tokens are. In the context of fine-level AudioLM, this suggests that the model does not necessarily need to synthesize the correct code to produce reasonable audio output. The bitrates are computed based on the future token's distributions obtained from LMCodec. For Huffman coding, we use the ground truth tokens encoded with the Huffman algorithm. Additionally, we note that the distributions of future tokens are updated every timestep based on the model, different from how other entropy codecs that may have fixed distributions operate. So, the Huffman bitrate may sometimes be lower than the bitrate derived from the entropy.

In this section, we additionally discuss some of the interesting audio effects from LMCodec. We suggest that readers listen to some of the audio samples from our model.
In particular, our model with only one quantizer is able to produce reasonable human voice with some babbling effects. The amount of babbling is reduced as the number of quantizers used in the codec increases. This suggests that there are some underlying hierarchical structure in SoundStream tokens, and the proposed codec can potentially be operating at very low bitrate, given that the coarse-to-fine prediction is accurate.

In Figure \ref{fig:audio-output}, we visualize the distribution of code prediction from the AudioLM model when the input is at the middle of a phoneme and between phonemes. We also found that the model is very confident if the audio input is the middle of the phonemes, as the language model network is able to learn underlying linguistic behavior of the utterances. On the other hand, the model has lower confidence in predicting the next token when reaching silence sections, suggesting that our proposed causal model is unable to predict future word really well. This confirms the babbling effect that we observed in the audio output from our proposed codec, which increases as we restrict the amount of information to describe each frame (e.g., by transmitting \linebreak fewer codes or dropping frames). 

Figure \ref{fig:mushra} shows the comparison of LMCodec with low-rate and medium-rate audio codecs. In particular, we find that LMCodec-4/6 performs better than SoundStream with 3 quantizers at \SI{1.5}{~}kbps but slightly worse than SoundStream with 12 quantizers at \SI{6}{~}kbps which is on par with LMCodec-8/12. We note that LMCodec-4/6 and LMCodec-8/12 are based on SoundStream with 6 and 12 quantizers respectively. Our results suggest that LMCodec effectively takes advantages from entropy coding and synthesizing reasonable fine-level codes from coarse-level codes. When comparing with SoundStream at similar rate, LMCodec essentially outperforms.

\subsection{Voice Activity Detection (VAD)}

\looseness -2
In this section, we show the performance of LMCodec applied only on audio regions with voice activity. We use an open-source RNNoise model \cite{valin2018hybrid}, which uses Mel-Frequency Cepstral Coefficients (MFCC) and outputs the probability of voice activity every \SI{10}{ms} frame size. Since the frame size of SoundStream tokens is \SI{20}{ms}, we run RNNoise on 2 consecutive 10-ms frames and define that the 20-ms SoundStream frame has a voice activity if and only if the probability that 2 consecutive frames have voice is over 0.8.

\looseness -2
Table \ref{tab:vad} shows the bitrate of LMCodec on two scenarios: (i) transmitting only voices and (ii) transmitting entire speech signals but using zero bits for non-voices. We report the  bitrate derived from the entropy and the bitrate based on Huffman coding. We note the first scenario has slightly lower bitrates as compared to bitrates from Table \ref{tab:accuracy} because the entropy for non-speech signals is usually higher than the entropy for speech signals. Additionally, the second scenario provides the lower bound estimate of bitrates when transmitting very low bits for non-voice signals similar to Opus with variable bitrate scheme.

\begin{table}
    \captionsetup{belowskip=-20pt}
    \centering
    \scriptsize
    \centerline{\begin{tabular}{lccc}
    \toprule
    $(N_\mathcal{C},N_\mathcal{F})$ & \textbf{Accuracy} & \textbf{Entropy}& \textbf{Huffman}\\
    \midrule\midrule
    % https://xm2a.corp.google.com/experiments/48089630
    $(2,1)$ & 15.5\% & \SI{534.0}{~}bps & \SI{542.5}{~}bps  \\
    % https://xm2a.corp.google.com/experiments/48088660
    $(3,1)$ & 14.3\% & \SI{837.1}{~}bps & \SI{845.7}{~}bps  \\
    % https://xm2a.corp.google.com/experiments/49313845
    $(4,2)$ & 13.1\% & \SI{1163.9}{~}bps & \SI{1173.5}{~}bps \\
    \midrule
    % https://xm2a.corp.google.com/experiments/49316968
    % new_scalable_12quantizers - 64 base_conv_depth
    $(1,11)$ & 16.1\% & \SI{262.8}{~}bps & \SI{262.6}{~}bps \\
    $(2,10)$ & 15.7\% & \SI{533.5}{~}bps & \SI{540.7}{~}bps \\
    $(3,9)$ & 14.9\% & \SI{844.6}{~}bps & \SI{847.4}{~}bps \\
    $(4,8)$ & 13.4\% & \SI{1154.2}{~}bps & \SI{1174.3}{~}bps \\
    $(6,6)$ & 11.9\% & \SI{1853.7}{~}bps & \SI{1861.2}{~}bps \\
    $(8,4)$ & 10.6\% & \SI{2561.8}{~}bps & \SI{2577.6}{~}bps \\
    $(10,2)$ & 9.7\% & \SI{3300.0}{~}bps & \SI{3324.8}{~}bps \\
    $(12,0)$ & 8.9\% & \SI{4094.5}{~}bps & \SI{4092.1}{~}bps \\
    \bottomrule
    \end{tabular}}
    \captionof{table}{Accuracy and bitrates. Bitrate without entropy coding is equivalent to \SI{500}{~}bps per quantizer (i.e., \SI{6}{~}kbps for 12 quantizers). Given the space limit, we only present the numerical results for LMCodec with 12 RVQ layers and LMCodec models shown in Figure \ref{fig:mushra}.
    \label{tab:accuracy}}
\end{table}

\begin{table}
    \captionsetup{belowskip=-13pt}
    \centering
    \setlength{\tabcolsep}{0pt} % make LaTeX figure out intercol. separation
    \scriptsize
    \begin{tabular}{@{\extracolsep{\fill}} l @{\extracolsep{0.5cm}}cccc}
         \toprule
         \multirow{2}{0.3cm}{$(N_\mathcal{C},N_\mathcal{F})$} & \multicolumn{2}{c}{Transmitting only voices} & \multicolumn{2}{c}{Transmitting non-voices with zero bits} \\
         \cmidrule(lr){2-3} \cmidrule(lr){4-5}
          & \textbf{Entropy} & \textbf{Huffman} & \textbf{Entropy} & \textbf{Huffman} \\
         \midrule \midrule
         $(2,1)$ & \SI{545.6}{~}bps & \SI{554.1}{~}bps  & \SI{303.1}{~}bps  & \SI{307.9}{~}bps \\
         $(3,1)$ & \SI{850.5}{~}bps & \SI{858.6}{~}bps  & \SI{472.1}{~}bps  & \SI{476.6}{~}bps \\
         $(4,2)$ & \SI{1165.6}{~}bps & \SI{1173.7}{~}bps & \SI{647.2}{~}bps & \SI{651.7}{~}bps  \\
        \midrule
         $(1,11)$ & \SI{268.3}{~}bps & \SI{268.7}{~}bps & \SI{149.3}{~}bps & \SI{149.5}{~}bps \\
         $(2,10)$ & \SI{523.7}{~}bps & \SI{530.0}{~}bps & \SI{290.8}{~}bps & \SI{294.3}{~}bps \\
         $(3,9)$ & \SI{816.5}{~}bps & \SI{819.1}{~}bps & \SI{453.2}{~}bps & \SI{454.7}{~}bps \\
         $(4,8)$ & \SI{1108.5}{~}bps & \SI{1129.7}{~}bps & \SI{615.2}{~}bps & \SI{627.0}{~}bps \\
         $(6,6)$ & \SI{1775.2}{~}bps & \SI{1783.6}{~}bps & \SI{985.5}{~}bps & \SI{990.2}{~}bps \\
         $(8,4)$ & \SI{2457.3}{~}bps & \SI{2471.5}{~}bps & \SI{1363.5}{~}bps & \SI{1371.4}{~}bps \\
         $(10,2)$ & \SI{3170.9}{~}bps & \SI{3196.7}{~}bps & \SI{1763.4}{~}bps & \SI{1777.8}{~}bps \\
         $(12,0)$ & \SI{3958.2}{~}bps & \SI{3951.5}{~}bps & \SI{2207.6}{~}bps & \SI{2203.9}{~}bps \\
         \bottomrule
    \end{tabular}
    \caption{Coding performance of LMCodec with VAD.}
    \label{tab:vad}
\end{table}

\subsection{Objective Evaluation}

\looseness -2
We present an objective evaluation on the audio examples from VCTK dataset \cite{Yamagishi2019CSTRVC} in Figure \ref{fig:objective_eval}. First, we demonstrate that the word error rate (WER) and character error rate (CER) are decreasing as the number of quantizers used in the LMCodec increases until around 4-6 quantizers, suggesting that the semantic content is stored in the coarse tokens. To evaluate WER and CER, we use two ASR models from AWS Transcribe service and Conformer model \cite{conformer} trained on LibriSpeech \cite{panayotov2015librispeech}. Second, ViSQOL \cite{chinen2020visqol} and WARP-Q \cite{jassim2021warp}, metrics designed for neural speech codecs, increases and decreases respectively, implying that the fine tokens are responsible for fine-grained acoustic details. Third, SSL-MOS  \cite{cooper2022generalization} shows that the overall \linebreak speech quality improves by increasing the number of quantizers.

\looseness -1
Despite neural speech codecs metrics ViSQOL and WARP-Q indicating worse performance at about 4-6 quantizers, our listening test shows very high quality audio results with small number of quantizers. This suggests that the language model of LMCodec is able to model the distribution of the fine tokens given the coarse tokens reasonably well even if the synthesized fine tokens are different from the ground truth ones. This drives metrics like ViSQOL and WARP-Q down as they primarily rely on the comparison between synthesized audio and its corresponding ground truth reference audio.

\looseness -1
When comparing LMCodec with different total number of quantizers, we first note that the upper bound performance of LMCodec with 6 quantizers is lower than the upper bound performance of LMCodec with 12 or 24 quantizers. However, LMCodec with a lower total number of quantizers reaches better performance faster than LMCodec with a higher total number of quantizers.

\section{Conclusion}
\label{sec:conclusion}

\looseness -2
Our experiments show that the proposed codec significantly outperforms the original neural speech codec with respect to the quality of synthesized speech when operating in the ultra-low bitrate regime. In addition, the subjective experiments indicate comparable to or better perceptual speech quality compared to conventional codecs operating at higher rates.
\clearpage
\bibliographystyle{IEEEbib}
\bibliography{refs}
\end{document}